\documentclass{pasj00}
\draft

\begin{document}
\SetRunningHead{Kominami and Makino}{Binary Formation in Planetesimal Disks II}

\title{Binary Formation in Planetesimal Disks II. Planetesimals with Mass Spectrum}

\author{Junko \textsc{D. Kominami} %
  \thanks{2-12-1-IE-1 Ohokayama, Meguro-ku, Tokyo, 152-8550, Japan}}
\affil{Earth-Life Science Institute, Tokyo Institute of Technology}
\email{kominami@mail.jmlab.jp}
\and
\author{Junichiro \textsc{Makino}}
\affil{RIKEN Advanced Institute for Computational Science\\
Earth-Life Science Institute, Tokyo Institute of Technology}\email{makino@mail.jmlab.jp}


%


\maketitle

\begin{abstract}
Many massive objects have been found in the outer region of the Solar
system.  How they were formed and evolved has not been well
understood, although there have been intensive studies on accretion
process of terrestrial planets.  One of the mysteries is the existence
of binary planetesimals with near-equal mass components and highly
eccentric orbits.  These binary planetesimals are quite different from
the satellites observed in the asteroid belt region.  The ratio of the
Hill radius to the physical radius of the planetesimals is much larger
for the outer region of the disk, compared to the inner region of the
disk.  The Hill radius increases with the semi major axis.  Therefore,
planetesimals in the outer region can form close and eccentric
binaries, while those in the inner region would simply collide.  In
this paper, we carried out $N$-body simulations in different regions
of the disk and studied if binaries form in the outer region of the
disk.  We found that large planetesimals tend to form binaries.  A
significant fraction of large planetesimals are components of the
binaries.  Planetesimals that become the components of binaries
eventually collide with a third body, through three-body encounters.
Thus, the existence of binaries can enhance the growth rate of
planetesimals in the Trans-Neptunian Object (TNO) region.
\end{abstract}

\section{Introduction}

In the trans-Neptuninan region, numerous planetesimal binaries have
been found.  So far, 79 planetesimal binaries have been found in the
Trans-Neptunian object (TNO) region
(http://www.johnstonsarchive.net/astro/tnoslist.html).  This number is
about $\sim10$\% of known TNOs between 30AU and 70AU (Stephens and
Noll, 2006), and $\sim 30$\% of the objects in the Cold Classical
Kuiper Belt (Noll et al., 2008a).

Some asteroids in the main belt have satellites (e.g. Chapman et
al. 1995).  These satellites have relatively small mass compared to
the primaries, and their eccentricities are quite small (Pravec et
al. 2006).  On the other hand, Trans-Neptunian binaries (TNBs) are
known to have almost equal brightness (Noll et al. 2008b), implying
comparable mass.  Hence, planetesimal binary systems with mass ratio
$\sim$1 are frequently observed in the TNO region.  The distribution
of the eccentricities of TNBs is pretty wide, from $\sim 0.1$ to $\sim
0.9$.

On the other hand, such systems have been hardly found in the asteroid
belt region.  The main physical difference between the inner region of
the disk and the outer region of the disk is the ratio between the
Hill radius and the physical radius of the planetesimals.  Hill radius
($r_{\rm h}$) of a planetesimal with mass $m$ and semi major axis of
$a$ is given by $(m/3M_\ast)^{1/3}a$.  Since it is proportional to
$a$, it is 30 times larger at 30AU compared to that of 1AU for a same
planetesimal.  Hence, the way two planetesimals interact can be quite
different in the TNO region and in the inner region, like the asteroid
belt region.  Some of the inner main belt objects are known to have
relatively small size ratio(around 1:3) and small separation (Johnston
2012).  They are most likely to be formed through the breakup of fast
rotating asteroids (Walsh \& Richardson 2006, Walsh et al. 2008,
2012).  One almost equal-sized binary system, 90 Antiope, was found in
2000 (Merline et al. 2000).  Its origin is so far not well understood,
but a recent observation indicates that two components are chemically
similar (Marchis et al. 2011), indicating the primordial origin of the
binary.

Since the physical ratio of planetesimals in unit of the Hill radius
are large in the inner region of the disk, two planetesimals can
easily collide.  Some of the collisions form fragments around
planetesimals, resulting in small planetesimals orbiting around large
planetesimal with small eccentricity.  On the other hand, in the TNO
region, Hill radii of the planetesimals are large.  Therefore,
physical collisions are less frequent, and complex interactions,
including three-body encounters, are more frequent compared to the
inner region.  Large planetesimals tend to have smaller random
velocities than those of small planetesimals due to the equipartition
of the energy.  Since the binaries are formed through three-body
encounters, the eccentricity of the binaries is not restricted to
small values.

Several formation scenarios have been proposed to explain the observed
characteristics of TNBs (Weidenschilling 2002, Goldreich et al. 2002,
Funato et al. 2004, Astakhov et al. 2005, Nesvorn${\rm \acute{y}}$ et
al. 2010).  Weidenschilling (2002) proposed a scenario in which two
planetesimals collide and accrete within the gravitational effect of
the third body.  However, collision frequency is too low to explain
the number of observed TNBs.  Goldreich et al. (2002) proposed two
mechanisms.  When the distance between two bodies becomes closer than
the Hill radius, a third planetesimal encounters with them and takes
away the energy leaving the two planetesimals being gravitationally
bound (Goldreich et al. 2002).  This is the scenario that we described
above.  Another mechanism Goldreich et al. (2002) proposed is that
instead of the third planetesimal, swarm of small planetesimals takes
away the energy of two encountering planetesimals (Goldreich et
al. 2002).  Exchange of a small satellite with another large
planetesimal is proposed by Funato et al. (2004).  Transitional binary
being tightened by several close encounters of other planetesimals
(Chaos-Assisted Capture) is proposed by Astackhov et al. (2005).
Planetesimal binary formation by gravitational instability is
introduced by Nesvorn${\rm \acute{y}}$ et al. (2010).

All of these studies, except for Nesvorn${\rm \acute{y}}$ et al., gave
purely theoretical models.  Nesvorn${\rm \acute{y}}$ et al. performed
the simulation of gravitational collapse of small planetesimals.
Except for this work, no self-consistent numerical study of formation
process of binaries in the TNO region is reported so far.  Kominami et
al. (2011) performed $N$-body simulation of disks of equal-mass
planetesimals at 30 AU in order to see if the binary formation is a
natural outcome in the TNO region.  They systematically changed
$r_{\rm h}/r_{\rm p}$ ($r_{\rm p}$ being the physical radius), and the
number density of planetesimals to study the effect of the binaries on
the collision rate of planetesimals.  They found that binaries are
involved in 1/3 - 1/2 of all collisions, and that the collision rate
is increased by about a factor of a few compared to the theoretical
estimate for the direct two-body collisions.  In the terrestrial
planet region, binaries are less important, because the ratio $r_{\rm
  h}/r_{\rm p}$ is relatively small.  Direct two-body collisions take
place instead of binary formation.  Although the duration of their
simulations was short, they clearly demonstrated that the accretion
process in the TNO region is different from that in the terrestrial
planet region.

In the present paper, we report the result of $N$-body simulations
with realistic mass distributions of planetesimals.  We show which
part of the mass spectrum is affected by the formation of the
binaries.  We found that the binary fraction of the massive
planetesimals is high, and that a significant fraction of collisions
between large planetesimals is through binary-single body
interactions.

In Section 2, we explain the calculation method.  Section 3 gives the
results.  We discuss the increase of the accretion rate due to the
binary formation in section 4.  The summary is given in section 5.

\newpage

\section{Calculation Model and Method}
\subsection{Initial Conditions}
We consider the initial model which is consistent with the standard
view of the collisional growth of the planetesimals.  The distribution
of planetesimal mass is a power-law and the random velocities of
planetesimals are in the thermal equilibrium.  We did not include the
effect of the gas drag for simplicity.  Also, we assumed the simple
perfect accretion model.  Thus, we effectively studied three-body
formation, dynamical-friction model, chaos-assisted capture, but not
models like Giant Impact or gravitational collapse.

Table 1 summarizes the initial conditions.  In all runs, the initial
models are narrow rings.  We studied models with four different values
of the initial semi-major axis (1,3,10 and 30).  The mass distribution
is given by
\begin{equation}
n {\rm d} m \propto m^{-p}{\rm d}m,
\end{equation}
where $n$ is the number density of planetesimal mass $m$.  The power
index $p$ is about 2.5 after runaway growth (Kokubo and Ida 1996).
Since the mass distribution of TNOs is not well understood, we adopt
the above power index ($p=2.5$) obtained in numerical simulations.
The minimum and maximum mass, $m_{\rm min}$ and $m_{\rm max}$, are
$2\times10^{22}$g and $2\times 10^{24}$g, respectively.  The values
$m_{\rm min}$ and $m_{\rm max}$ have been fixed for simplicity.  The
observed TNOs are about $\sim 100$ km in size.  In order to keep the
simulations feasible, we constrained the number of planetesimals and
hence have to keep the planetesimal mass large.  From theoretical
point of view, growth timescale and other behavior would be the same
if physical quantities are properly scaled with the Hill radius.  So
we believe the relatively large mass used in our study does not cause
serious problems.  In order for the ring not to be too narrow, we set
the planetesimals to be large as described.  The number density used
here is the value derived from the so-called minimum-mass solar nebula
(Hayashi 1981).  The eccentricity and inclination of planetesimals are
given by the Maxwellian distribution.  The initial RMS eccentricity of
planetesimals is given by
\begin{equation}
\langle e^2 \rangle^{1/2} = \eta \left(\frac{m_{\rm
    max}}{3M_\odot}\right)^{1/3}\left(\frac{m}{m_{\rm
    max}}\right)^{-1/2},
\label{eq.ecc}
\end{equation}
where $\eta$ is a parameter.  We varied $\eta$ from 0.05 to 0.4 in
order to study the effect of the initial velocity dispersion.  For all
models, we used $\langle e^{2} \rangle^{1/2}=2 \langle i^{2}
\rangle^{1/2}$ (Ida and Makino, 1992).  We create initial models so
that there is no binary.

\subsection{Orbital Integration}
The equation of motion of a planetesimal is
\begin{equation}
\frac{{\rm d}{v}_j}{{\rm d}t}=-\frac{GM_\odot}{|{r}_j|^3}
{r}_j - \sum_{k \neq j} \frac{Gm_{j}}
{|{r}_j - {r}_k|^3}
\left( {r}_j - {r}_k \right), 
\label{eq.motion}
\end{equation}
where ${r}_j$ is the heliocentric position vector, ${v}_j$ is the
velocity vector, and $m_{j}$ is the mass of particle $j$, and $G$ and
$M_\odot$ are the gravitational constant and the solar mass,
respectively.  The first and second terms are the gravity of the Sun
and the mutual gravity between the particles, respectively.  Since the
total mass of the planetesimals is $\sim 10^{-6} M_\odot$, we neglect
the indirect term.  We use the 4th-order Hermite scheme (Makino and
Aarseth 1992, Kokubo et al. 1998) using individual time step with
block step algorithm for orbital integration.  The energy error was
monitored in the simulations.  The error was within $10^{-6}$ level
throughout the simulations.  The mutual gravity term is calculated
using GRAPE-DR (Makino et al. 2007).  The integration time is $10^4$
years.  The number of binaries becomes roughly constant after several
thousand years.

We take into account the particle accretion following the treatment in
the previous works (e.g., Kokubo and Ida 1996,1998).  For simplicity,
we use the perfect accretion model, in which we let two planetesimals
merge when the distance between them becomes less than the sum of
their radii.  When the binary induced collisions take place, the
relative velocity of the components can be high enough to produce some
fragmentation.  This effect has not been considered yet, and should be
discussed in future study.  The physical radius of a planetesimal is
determined by its mass $m$ and the density $\rho$ as
\begin{equation}
R = \left( \frac{3}{4 \pi} 
	\frac{m}{\rho}\right)^{1/3}.
\label{hankei}
\end{equation}
We assume $\rho = 3 {\rm g}$ ${\rm cm}^{-3}$.  The solid density of
the planetesimals in the TNO region is not well understood, and it
might be more reasonable to use a smaller value.  We adopted the
typical density in the terrestrial region, which is not far from that
of the observational value ($\sim 1-2{\rm g}$ ${\rm cm}^{-3}$,
e.g. Grundy et al. 2007, Johnston 2012).

\subsection{Binary Definition}
In this paper, we define a pair of planetesimals as binary, if their specific 
Jacobi energy $E_{\rm J}$, defined as,
\begin{equation}
E_{\rm J}=\frac{1}{2}(\dot{x}^2+\dot{y}^2+\dot{z}^2)
-\frac{3}{2}\Omega_{\rm K}^2x^2 + \frac{1}{2}\Omega_{\rm K}^2z^2
-\frac{G(m_{1}+m_{2})}{r} + \frac{9}{2}r_{\rm H}^2\Omega_{\rm K}^2,
\end{equation}
under the Hill approximation (Nakazawa and Ida 1988), is negative,
where $x,y,z$ are relative cartesian coordinates of two planetesimals
whose masses are $m_{1}$ and $m_{2}$ and $\Omega_{\rm K}$ is the
Kepler angular velocity at the barycenter given as
$\sqrt{GM_\odot/a^3}$.  When the distance between two planetesimals
becomes smaller than the mutual Hill radius,$r_{\rm H}$, we calculate
$E_{\rm J}$.  The mutual Hill radius $r_{\rm H}$ of $m_1$ and $m_2$ is
defined as
\begin{equation}
r_{\rm H} = \left( \frac{m_1+m_2}{3M_\odot} \right)^{1/3}
\left( \frac{a_1m_1+a_2m_2}{m_1+m_2} \right),
\end{equation}
where $a_1$ and $a_2$ are the semi major axes of $m_1$ and $m_2$,
respectively.

There are some pairs of which components orbit around each other but
have positive $E_{\rm J}$.  However, the number of such pairs is
negligible and they do not make substantial difference in the results.


\section{Results}
\subsection{Evolution of Velocity Dispersion}
\begin{figure}
\begin{center}
\includegraphics[scale=0.7]{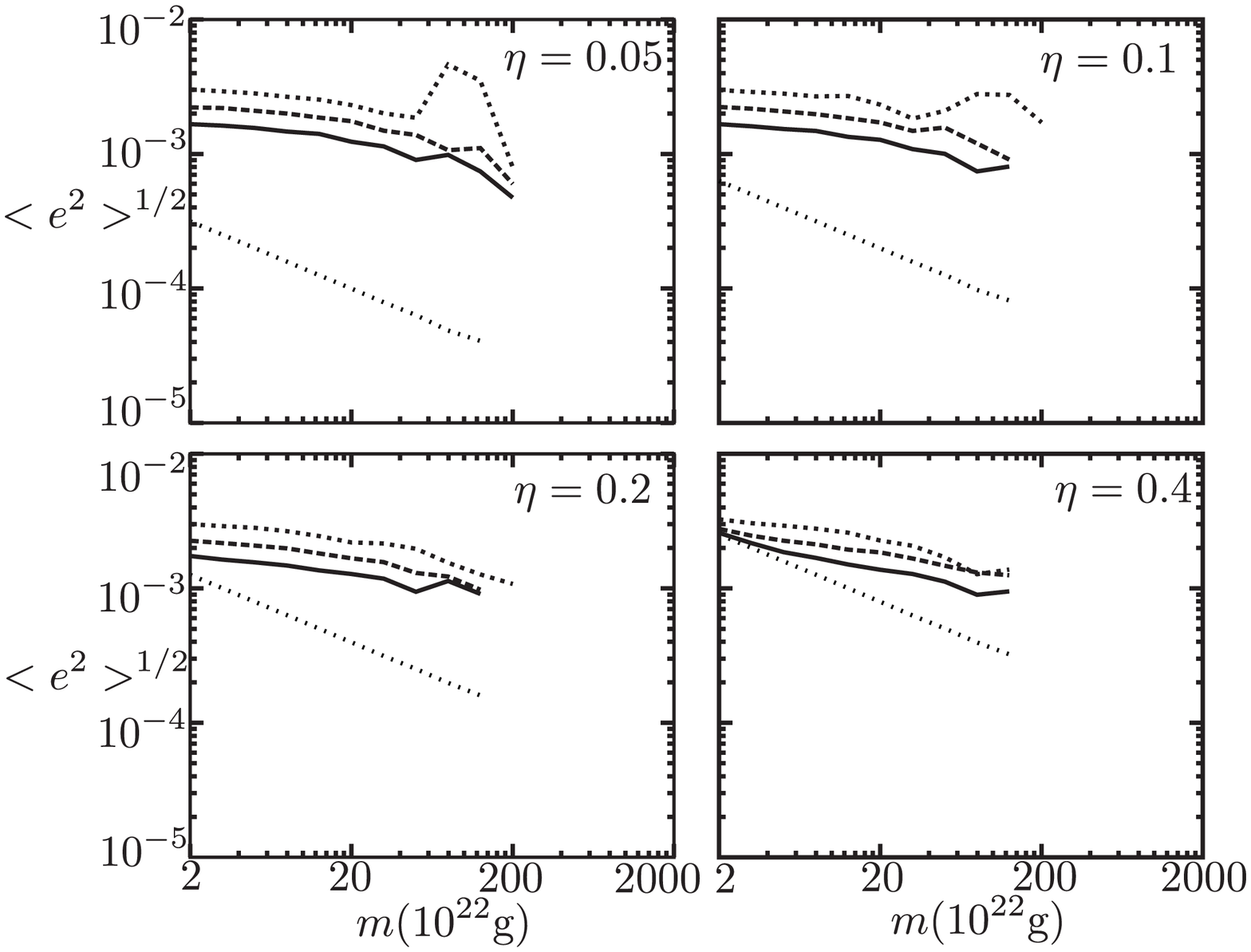}
\end{center}
\caption{RMS eccentricity plotted as a function of mass for runs
S30e0.05, S30e0.10, S30e0.20 and S30e0.40.
The solid line is for $1000$ years, the long dashed line is for $3000$ years,
the short dashed line is for $10^4$ years and the dotted line is for the value of
initial condition, respectively.}
\label{fig.30AU_mass_ecc.new}
\end{figure}

\begin{figure}
\begin{center}
\includegraphics[scale=0.7]{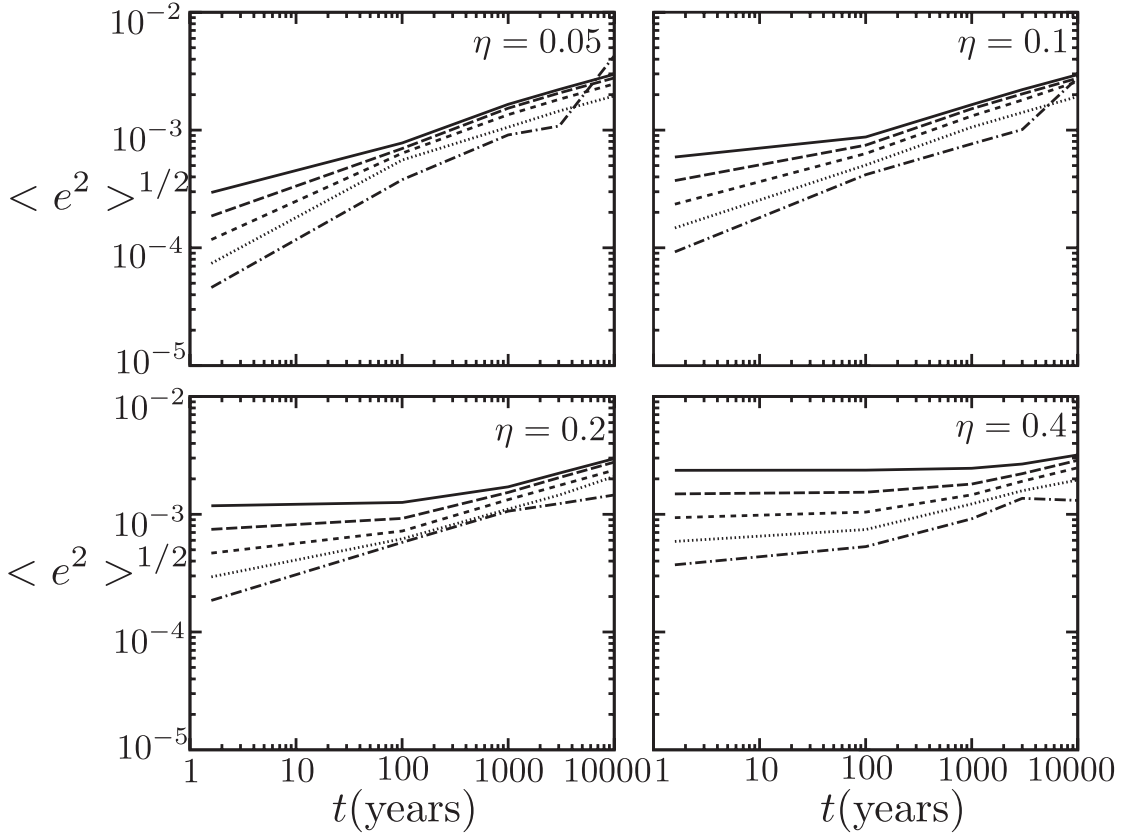}
\end{center}
\caption{Time evolution of the RMS eccentricity in five mass bins.
Solid curve corresponds to the mass range
$2.0\times 10^{22}$g - $5.0\times 10^{22}$g,
long dashed curve corresponds to the mass range
$5.0\times 10^{22}$g - $1.26\times 10^{23}$g,
short dashed curve corresponds to the mass range
$1.26\times 10^{23}$g - $3.17\times 10^{23}$g,
dotted curve corresponds to the mass range
$3.17\times 10^{23}$g - $7.96\times 10^{23}$g ,
and dot-dashed curve corresponds to the mass range
$7.96\times 10^{23}$g - $2.0\times 10^{24}$g, respectively.
The results are for runs S30e0.05, S30e0.10, S30e0.20 and S30e0.40.}
\label{fig.MBIN_30AU.new}
\end{figure}

Figure \ref{fig.30AU_mass_ecc.new} shows the evolution of the velocity
dispersion for four runs: S30e0.05, S30e0.1, S30e0.2 and S30e0.4.  The
plot with $\eta=0.05$ corresponds to the simulation result of
S30e0.05, $\eta=0.1$ corresponds to S30e0.1, $\eta=0.2$ corresponds to
S30e0.2 and $\eta=0.4$ corresponds to S30e0.4.  We can see that the
velocity dispersion after $10^3$-$10^4$ years is pretty similar for
all runs.  The velocity dispersion is almost flat for $m < 2\times
10^{23}$g, then approaches to the thermal equilibrium of $\langle e^2
\rangle \propto m^{-1}$ for larger mass.  Figure
\ref{fig.MBIN_30AU.new} shows the time evolution of the mean
eccentricities for models S30e0.05, S30e0.10, S30e0.20 and S30e0.40.
The mass range is divided into 5 bins with the same logarithmic width.
The evolution of the eccentricity is quite similar for different mass
bins after 100 - 1000 years, and reaches to similar values independent
of initial values of $\eta$ after $10^4$ years.  Note that since the
number of the planetesimals in the largest mass bin is small, there
are some fluctuations, especially when $\eta$ is small.  The velocity
dispersion is far from the thermal equilibrium for the low-mass part.
This is because the equipartition time scale is much longer than the
heating time scale.  On the other hand, the high-mass end ($m>5\times
10^{23}$g) is approximately in the equipartition, which means massive
planetesimals have smaller random velocity.

\subsection{Evolution of Mass Distribution}
\begin{figure}
\begin{center}
\includegraphics[scale=0.7]{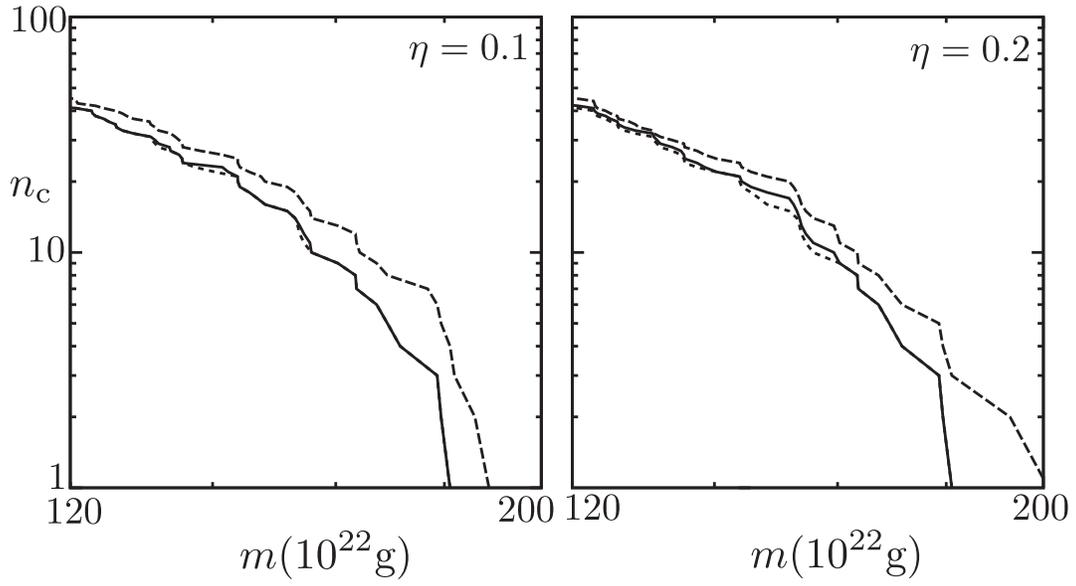}
\end{center}
\caption{Cumulative mass distribution of planetesimals 
of runs S30e0.10 and S30e0.20.
Solid, long dashed and short dashed curves show the distributions at
$t=10^3, 10^4$ years, and the initial mass distribution, respectively.}
\label{fig.mass.eta0.1.0.2.new}
\end{figure}

Figure \ref{fig.mass.eta0.1.0.2.new} shows the mass distributions at
$t=10^3, 10^4$ years and initial mass distribution for runs S30e0.10
and S30e0.20, which are the realistic models in the sense that initial
velocity dispersion is fairly high.  When we compare the long dashed
curve and short dashed curve, it is clear that the number of
planetesimals with mass greater than $\sim 2\times 10^{24}$g increases
with time.  As we will see in section 3.5, most of the collisions
after $10^3$ years are binary induced collisions.

\subsection{Binary Formation Process}
\begin{figure}
\begin{center}
\includegraphics[scale=0.7]{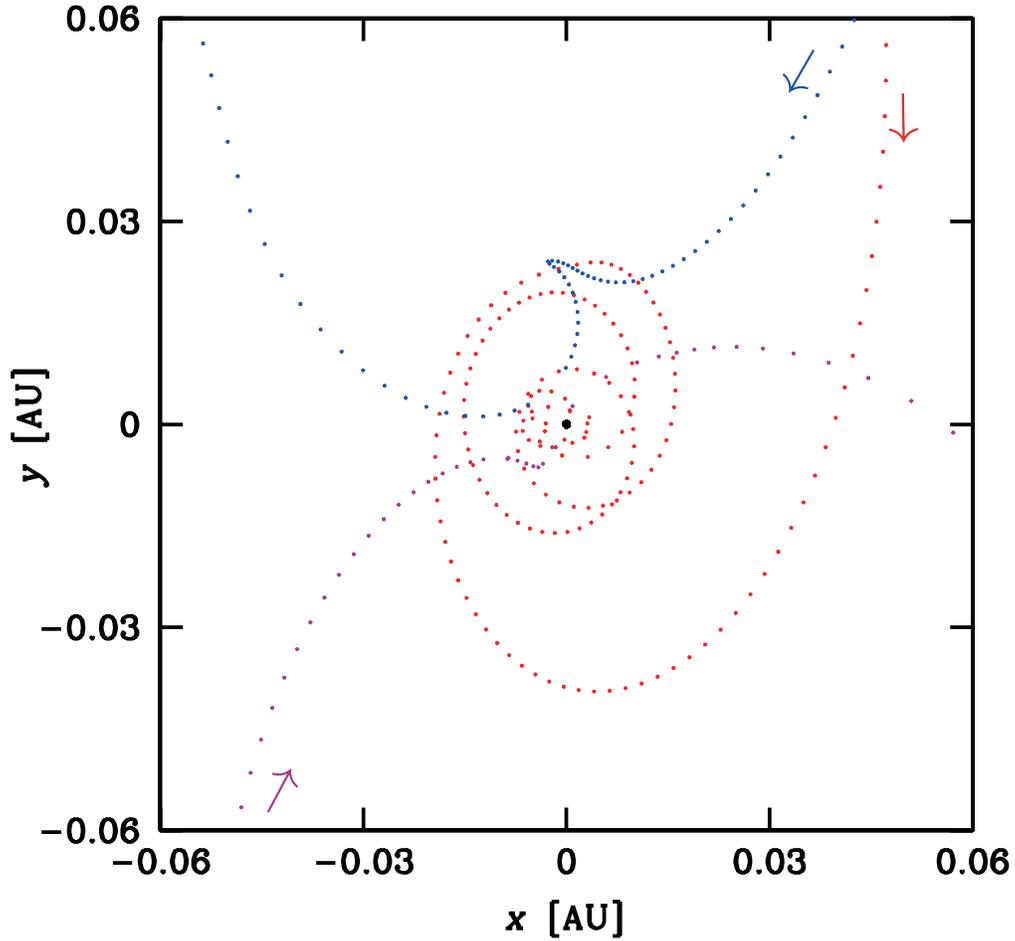}
\end{center}
\caption{Trajectories of three planetesimals during a binary formation event.
Origin is placed at position of one of the final binary
components (black dot).   Each dot is plotted in rotational frame around
the black one.
First, the planetesimal colored in blue encounters the planetesimal
colored in red.  It makes the red planetesimal temporally bound to the
black planetesimal.
Then the magenta planetesimal approaches the red planetesimal
taking away energy enough to make the red planetesimal bound stably.}
\label{fig.BIN1712}
\end{figure}

\begin{figure}
\begin{center}
\includegraphics[scale=0.7]{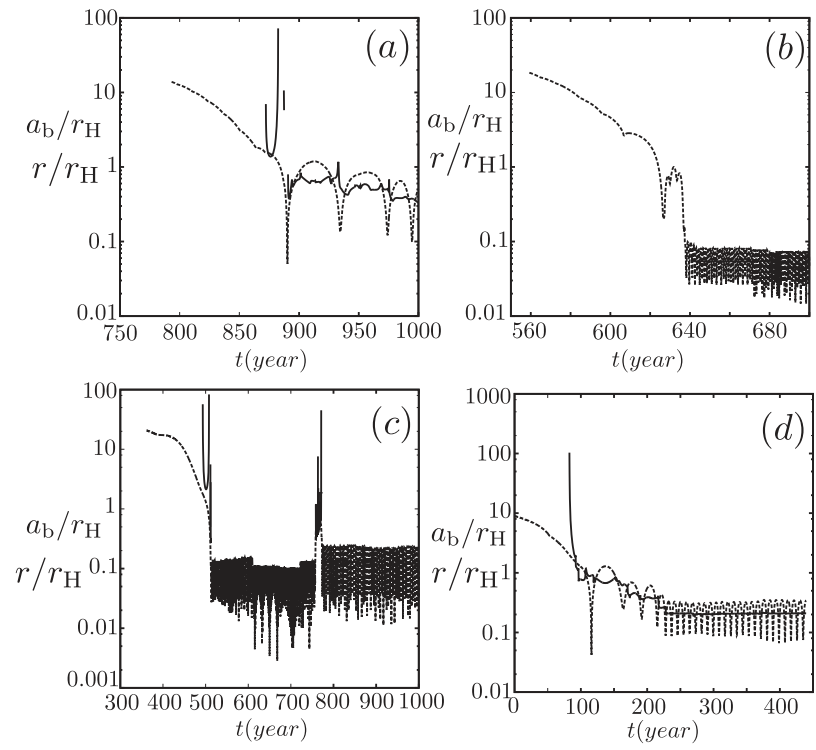}
\end{center}
\caption{The time evolution of the semi major axis $a_{\rm b}$
and separation $r$ of two components of a binary in unit of the
Hill radius $r_{\rm H}$.
The solid and dashed curves indicate $a_{\rm b}$ and $r$.}
\label{fig.BIN1-4.new2}
\end{figure}

Figure \ref{fig.BIN1712} shows an example of the binary formation
process in run S30e0.05.  First, blue and red dots approach to each
other, and experience a close encounter.  After the encounter, the red
body becomes weakly bound to the black one.  Then third planetesimal
(magenta) encounters with the red body and takes away some energy.  As
a result, the red body becomes bound to the black body.  This
formation process is the three-body process discussed in Goldreich et
al. (2002).  Figure \ref{fig.BIN1-4.new2} shows examples of the binary
formation process.  The dotted curves show the distance between the
binary components and the solid curves show the semi major axis.  The
quantities are plotted when Jacobi energy is negative.  In all of
these cases, a third body is involved in the formation of the binary.

\subsection{Mass Distribution of Binaries}
\begin{figure}
\begin{center}
\includegraphics[scale=0.7]{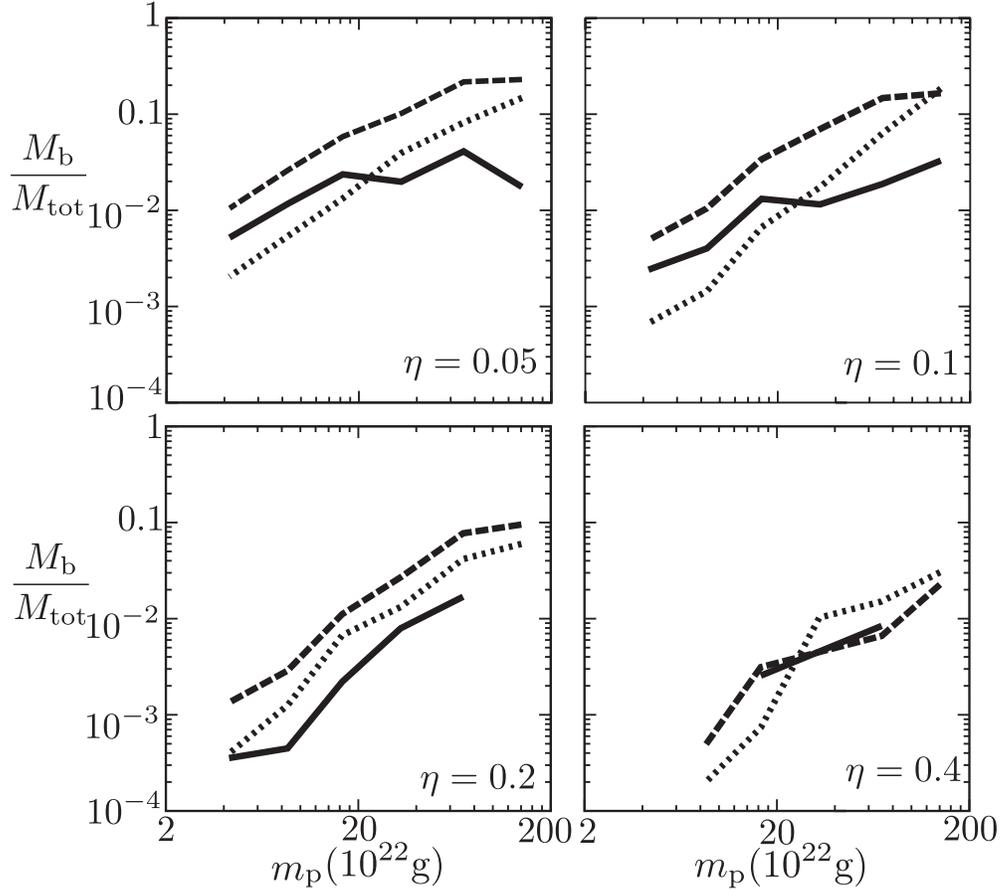}
\end{center}
\caption{Mass fraction of binaries as the function of the mass of the primary
for runs S30e0.05, S30e0.10, S30e0.20 and S30e0.40.
Solid, long dashed and short dashed curves corresponds to 
$t=10^2,10^3$ and $10^4$ years, respectively.}
\label{fig.bin_mass_bin_new}
\end{figure}

\begin{figure}
\begin{center}
\includegraphics[scale=0.7]{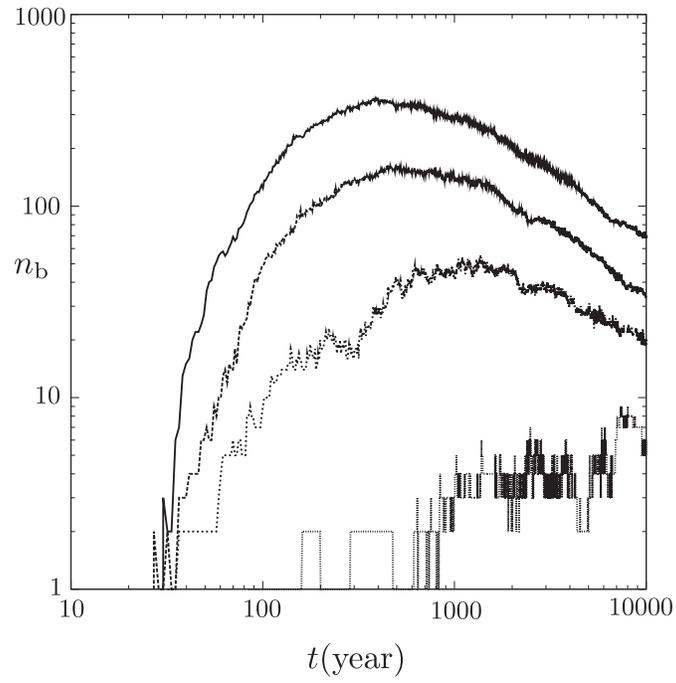}
\end{center}
\caption{Time evolution of number of binaries for runs
S30e0.05(Solid), S30e0.10 (long dashed), S30e0.20 (short dashed) and S30e0.40 (dotted).}
\label{fig.NB_time_30AU.new}
\end{figure}

\begin{figure}
\begin{center}
\includegraphics[scale=0.7]{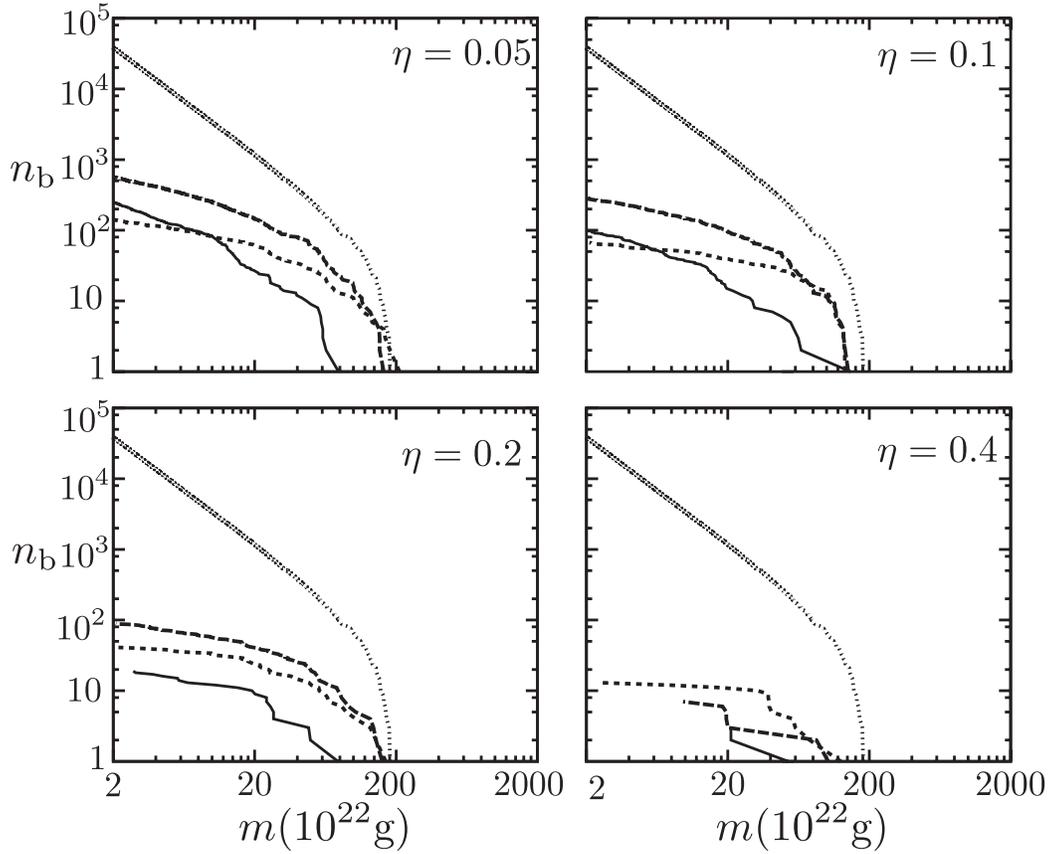}
\end{center}
\caption{Cumulative mass distribution of the binary components for runs
S30e0.05, S30e0.10, S30e0.20 and S30e0.40.
Both components are counted.
Solid, long dashed, short dashed and dotted curves show the distributions at
$t=10^2,10^3$ $10^4$ years, and the initial mass distribution, respectively.}
\label{fig.BIN_mass_30AU.new}
\end{figure}

Figure \ref{fig.bin_mass_bin_new} shows the fraction of binary mass in
each mass bin.  The solid line corresponds to $10^2$ years, long
dashed line to $10^3$ years and the short dashed line to $10^4$ years.
The difference is not so large, but if we compare the curves at $10^3$
and $10^4$ years, we can see that binary fraction shows some decrease
in the low-mass side but essentially unchanged in the high-mass side.
Figure \ref{fig.NB_time_30AU.new} shows the time evolution of the
number of binaries in runs S30e0.05, S30e0.1, S30e0.2 and S30e0.4.  In
the runs with small $\eta$, many binaries are formed initially, when
the velocity dispersion is low.  In the case of run with $\eta=0.4$,
the total number of binaries keeps increasing, even though the
velocity dispersion is increasing slowly.  Initial increase of the
binaries is the effect of the initial condition.  Binaries form more
easily when the random velocities of the planetesimals are small.  The
increase in the random velocity is very large for models with low
initial eccentricity.  In these models many very soft binaries are
initially formed, and most of them were destroyed after the random
velocity increased.  In figure \ref{fig.BIN_mass_30AU.new}, we can see
that the large fraction of binary components are massive ($m>5\times
10^{22}$g) at $t=10^{4}$ years.  At $t=10^4$ years, $\sim$ 10\% of
planetesimals in the highest mass bin are in binaries for all runs.
This result shows that large planetesimals tend to form binaries, and
once they are formed, they stay as binaries.  In the runaway growth
model, the behavior of the massive bodies determine the evolution.
Since large planetesimals tend to form binaries, the accretion process
can be significantly affected by the presence of binaries.

\subsection{Number of Collisions}
Although the number of collisions in the outer region is fewer than
that in the inner region, a significant number of planetesimals
experience collisions in runs S30e0.05, S30e0.10, S30e0.20 and
S30e0.40.  There are two types of collisions.  The first is the
ordinary collision which is of the same type of as what happens in the
inner region.  Another type of collision is the binary induced
collision.  First, two bodies become gravitationally bound through
three-body interaction.  Then a third body perturbs their orbits and
one of the components collide with the third body or with the other
component.  The number of such collisions is 103 in S30e0.05, 71 in
S30e0.10, 26 in S30e0.20 and 11 in S30e0.40.

In order to determine the type of a collision, we carried out the
following procedure.  We look at the snapshot $100/(2\pi)$ years
before the collision.  First we check the Jacobi energy of the
colliding two bodies and see if it is negative.  If so, we count them
as a binary.  If not, we look for the neighboring particles of the
colliding particles.  If we find a third body within the mutual Hill
radius of the colliding particles, we calculate the Jacobi energy of
the closest two bodies (one of the colliding two and the third body).
If the Jacobi energy is negative, that collision is also counted as
binary-induced collision.  The Jacobi energy of the two particles is
calculated using the position and the velocity vector.  We checked if
the colliding two bodies are really orbiting around each other at
least several orbits.

\begin{figure}
\begin{center}
\includegraphics[scale=0.7]{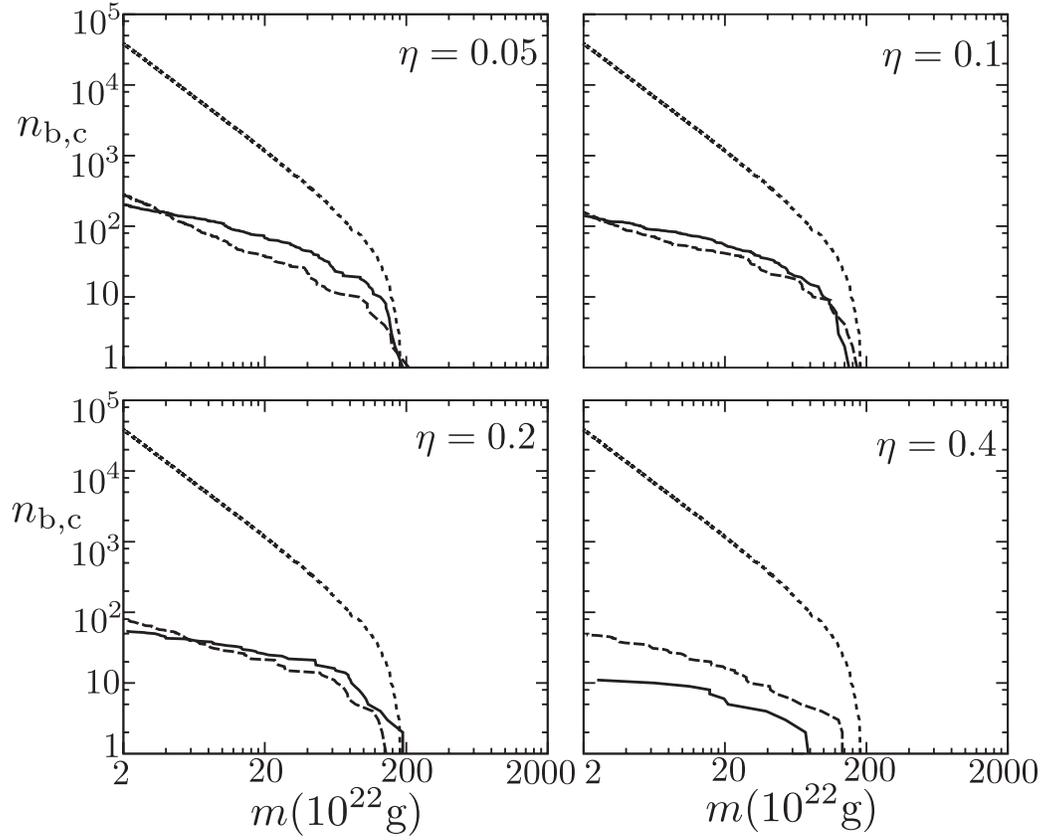}
\end{center}
\caption{Cumulative mass distribution of planetesimals that experienced binary induced
collision(solid curve) and non-binary-induced ordinary collision(long dashed curve) 
for runs S30e0.05, S30e0.10, S30e0.20 and S30e0.40.
Initial mass distribution is shown for comparison in short dashed curve}
\label{fig.COLL_tmp2_alleta.new}
\end{figure}

Figure \ref{fig.COLL_tmp2_alleta.new} shows the mass distribution of
particles participated in collisions in runs S30e0.05, S30e0.10,
S30e0.20 and S30e0.40.  Solid and dashed curves indicate
binary-induced and non-binary-induced collisions, respectively.  The
slope of the mass distribution of collided planetesimals is shallower
than that of the initial mass distribution.  Moreover, the slopes of
the solid curves are somewhat shallower than that of the dashed
curves, which means that the fraction of binary induced collisions
over the total number of collisions is higher for larger
planetesimals.

\begin{figure}
\begin{center}
\includegraphics[scale=0.7]{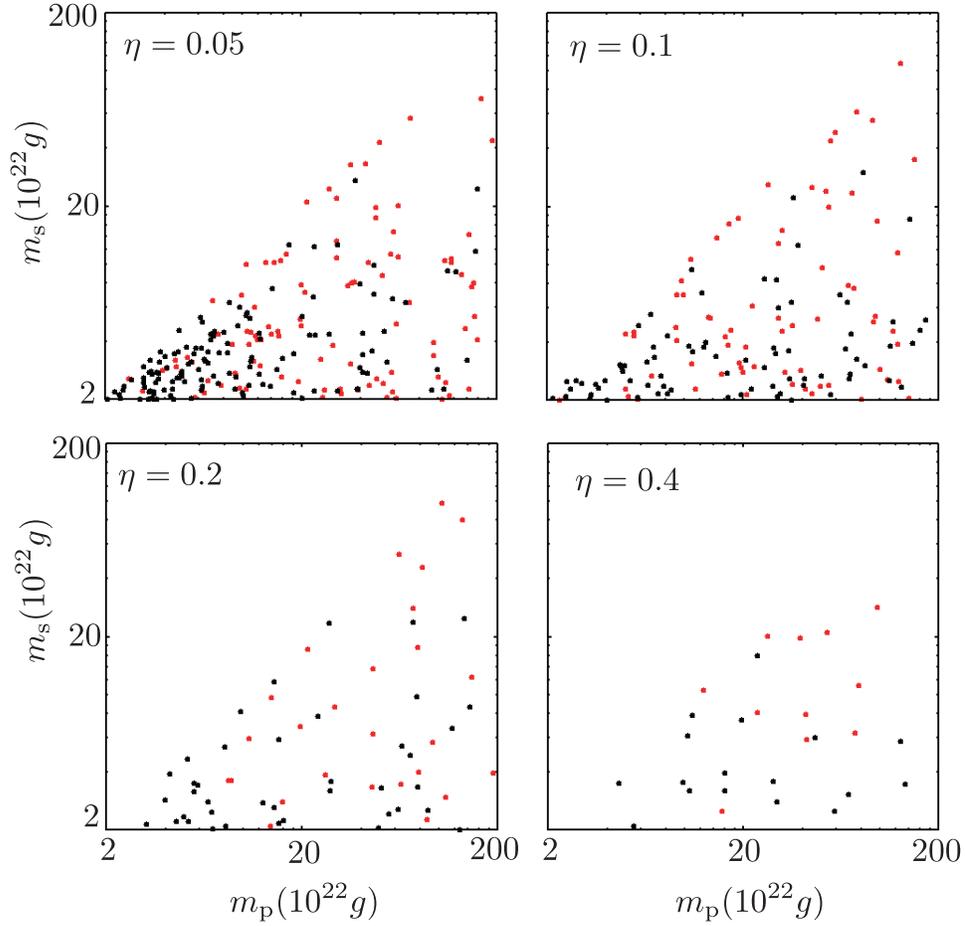}
\end{center}
\caption{Primary mass ($m_{\rm p}$) and secondary mass ($m_{\rm s}$) of all collisions
for runs S30e0.05, S30e0.10, S30e0.20 and S30e0.40.
Binary induced collisions are plotted in red and non-binary-induced
collisions in black.}
\label{fig.MassDistAllEta2.new2}
\end{figure}

In figure \ref{fig.MassDistAllEta2.new2}, we show all collisions in
runs S30e0.05, S30e0.10, S30e0.20 and S30e0.40 on the $m_{\rm
  p}$-$m_{\rm s}$ plane, where $m_{\rm p}$ and $m_{\rm s}$ are the
mass of particles collided ($m_{\rm p}\geq m_{\rm s}$).  It is clear
that binary-induced collisions have systematically larger values of
$m_{\rm s}/m_{\rm p}$ compared to non-binary-induced collisions.  To
see this tendency more quantitatively, we calculated average values of
$m_{\rm s}/m_{\rm p}$ for massive and less massive primaries.

\begin{figure}
\begin{center}
\includegraphics[scale=0.7]{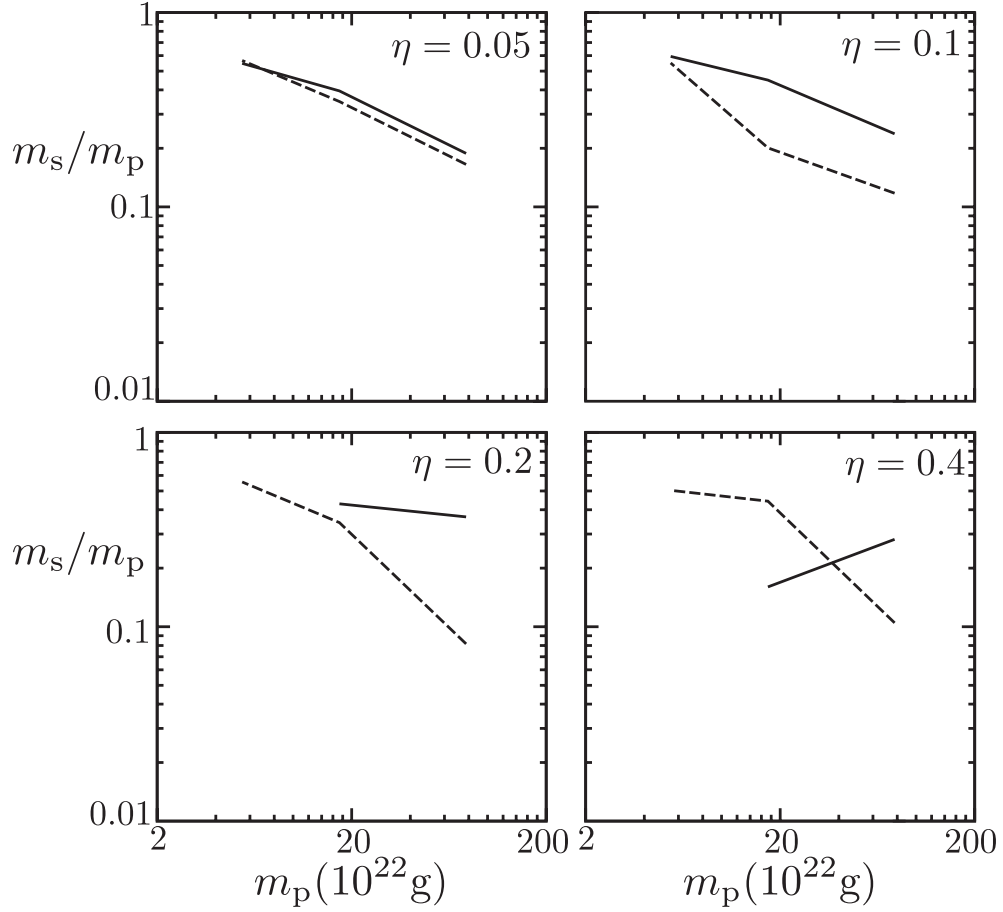}
\end{center}
\caption{Mass ratio $m_{\rm s}/m_{\rm p}$ of binary-induced collisions(solid line)
and of non-binary-induced collisions(long dashed line) as a function of 
primary mass $m_{\rm p}$
for runs S30e0.05, S30e0.10, S30e0.20 and S30e0.40.}
\label{fig.COLL6_mass_bin}
\end{figure}

The result is shown in figure \ref{fig.COLL6_mass_bin}.  In the case
of $\eta=0.4$, $m_{\rm s}/m_{\rm p}$ is higher for larger $m_{\rm p}$
in the case of binary-induced collisions, while it is lower for larger
$m_{\rm p}$ in the case of non-binary-induced collisions.  Though not
this clear, similar tendency is visible for other values of $\eta$,
except for $\eta=0.05$ where the difference is small.  Since the
number of collisions is similar for two types of collisions, this
difference in the average mass of the secondary means that the growth
of the mass of massive bodies comes primarily from binary-induced
collisions.  In other words, massive planetesimals in the TNO region
grow through binary-induced collisions, not through ordinary,
non-binary induced collisions.

\begin{figure}
\begin{center}
\includegraphics[scale=0.7]{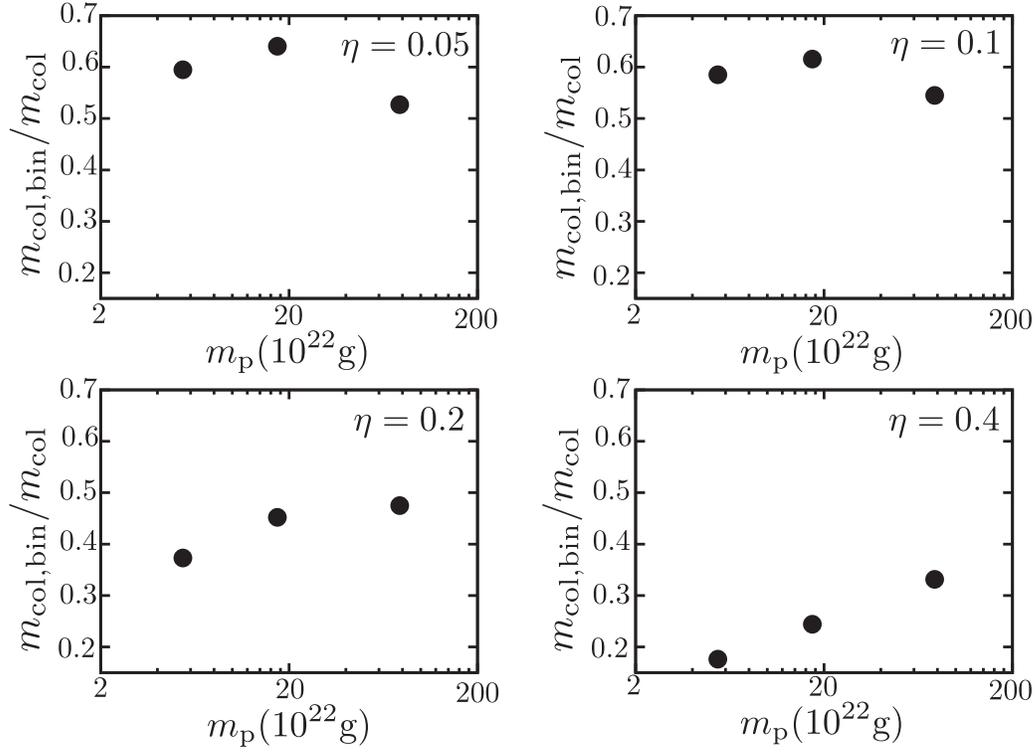}
\end{center}
\caption{The total mass of binary-induced colliding pairs normalized by total mass of
planetesimals experienced collision, plotted as a function of the primary mass.
The results are from runs S30e0.05, S30e0.1, S30e0.2 and S30e0.4.}
\label{fig.mass_bin_col}
\end{figure}

Figure \ref{fig.mass_bin_col} shows the mass fraction of the
binary-induced collisions over total collisions, as the function of
the primary mass.  For runs with large $\eta$(0.2 and 0.4), we can see
the tendency that the fraction of binary-induced collision is higher
for larger $m_{\rm p}$.

We can conclude that, for massive planetesimals, binary-induced
collisions are frequent, and they are the main route of the growth of
massive planetesimals.


\section{Discussion}
Here we estimate the collision probability increase due to the
formation of planetesimal binaries.  Let the binary induced collision
probability to be $P_{\rm c,bin}$, usual collision probability $P$ and
the total collision probability be $P_{\rm total}= P_{\rm c,bin}+P$.
From Makino et al. (1998), usual collision probability of $m$ and $m'$
is
\begin{equation}
P = \frac{1}{\max(H,H')}\pi (r+r')^2\left(1+\frac{v_{\rm esc}^2}{v_{\rm rel}^2}
\right)v_{\rm rel}
\label{eq.Pcol.ord}
\end{equation}
where $H$ is the scale height, $r$ is the physical radius of mass $m$,
$v_{\rm esc}= \sqrt{2G(m+m')/(r+r')}$ is the escape velocity and
$v_{\rm rel}^2=v^2+v'^2$ is the relative velocity.  The value $v_{\rm
  rel}$ can be written as
$v_{\rm rel}\sim e v_{\rm k} = e \sqrt{G(m+m')/r_{\rm c}}$, 
where $r_{\rm c}=(mr+m'r')/(m+m')$ is $r$ of the center of mass of $m$ and $m'$.
Hence, $P$ can be written as 
\begin{equation}
P = \frac{1}{\max(H,H')}\pi (r+r')^2\left(1+\frac{2}{e^2} \frac{mr+m'r'}{(r+r')(m+m')}
\right)e\sqrt{\frac{G(m+m')^2}{mr+m'r'}}
\label{eq.Pcol.ord2}
\end{equation}
On the other hand, binary induced collision probability 
$P_{\rm c,bin}$ can be written as
\begin{equation}
P_{\rm c,bin} = \frac{P_{\rm b}}{\max(H,H')}\pi a_{\rm b}^2P_{\rm L3}
\left(1+\frac{v_{\rm esc,b}^2}{v_{\rm rel}^2}
\right)v_{\rm rel}
\label{eq.PcolBIN}
\end{equation}
where 
\begin{equation}
v_{\rm esc,b} = \sqrt{\frac{2G(m+m')}{a_{\rm b}}}
\end{equation}
and $a_{\rm b}$ is the typical semi major axis of binaries.
We assume $a_{\rm b}\sim r_{\rm H}/10$.
Here, $P_{\rm L3}$ is the probability of physical collision after a 
binary experienced close encounter with a third body.
Eq.(\ref{eq.PcolBIN}) can be written as 
\begin{equation}
P_{\rm c,bin} = \frac{P_{\rm b}}{\max(H,H')}\pi a_{\rm b}^2 P_{\rm L3}
\left(1+\frac{2(mr+m'r')}{e^2a_b(m+m')}
\right)e \sqrt{\frac{G(m+m')^2}{mr+m'r'}}.
\label{eq.Pcol.BIN2}
\end{equation}
Hence, from (\ref{eq.Pcol.BIN2}) and (\ref{eq.Pcol.ord2}) we obtain 
\begin{equation}
%
\frac{P_{\rm c,bin}}{P} = P_{\rm b} \frac{a_{\rm b}^2}{(r+r')^2}
\left( \frac{1+2(mr+m'r')/(e^2 a_{\rm b}(m+m'))}{1+2(mr+m'r')/(e^2(r+r')(m+m'))} \right)P_{\rm L3}
\label{Pratio}
\end{equation}
Let us assume $a_{\rm b}\sim r_{\rm H}/10$, $r_{\rm H}\sim 100r$.
Also, if we assume $1\ll 2 (mr+m'r')/(e^2a_{\rm b}(m+m'))$ and $1\ll 2(mr+m'r')/(e^2(r+r')(m+m'))$, 
eq(\ref{Pratio}) becomes
\begin{equation}
\frac{P_{\rm c,bin}}{P}\sim P_{\rm b} \frac{a_{\rm b}^2}{(r+r')^2}
\frac{r+r'}{a_{\rm b}}P_{\rm L3}\sim P_{\rm b} P_{\rm L3}\frac{a_{\rm b}}{r+r'}
\end{equation}
As shown in the previous section, When $m$ is large, $P_{\rm b}\sim 0.1$.  
If we assume $P_{\rm L3}\sim 1$, we have
\begin{equation}
P_{\rm total} = P_{\rm c,bin} + P = 
(1+P_{\rm b}P_{\rm L3}\frac{a_{\rm b}}{r+r'}) P\sim P.
\end{equation}
Thus, roughly speaking, the rate of binary-induced collision is
comparable to that of non-binary induced collisions, in our model
calculations.  The enhancement of the growth rate due to the
binary-induced collisions is larger for the following two reasons.

First, as discussed in section 3.4, the secondary mass is larger for
the binary-induced collisions by a factor of three or around.  This
difference directly results in the increase of the growth rate by the
same factor.

\begin{figure}
\begin{center}
\includegraphics[scale=0.7]{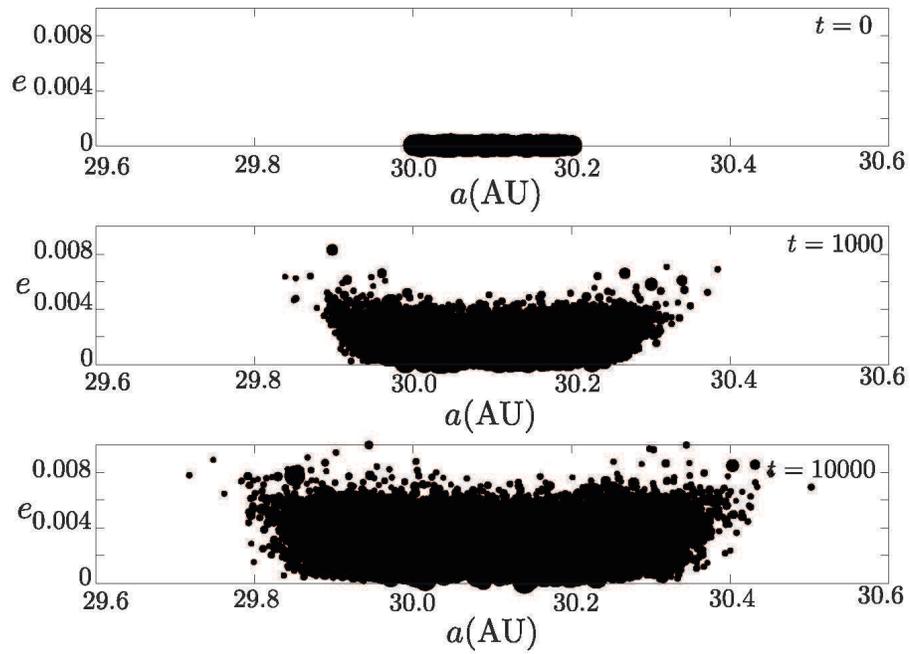}
\end{center}
\caption{Distribution of planetesimals in the plane
of semi-major axis and the eccentricity.  Top, middle and bottom
panels show the distribution at $t=0, 1000$ and 10000 years, 
respectively.  The radius of the particles corresponds to the size of the planetesimals.
The initial model is S30e0.05.}
\label{fig.run1614_new_eta0.05}
\end{figure}

\begin{figure}
\begin{center}
\includegraphics[scale=0.7]{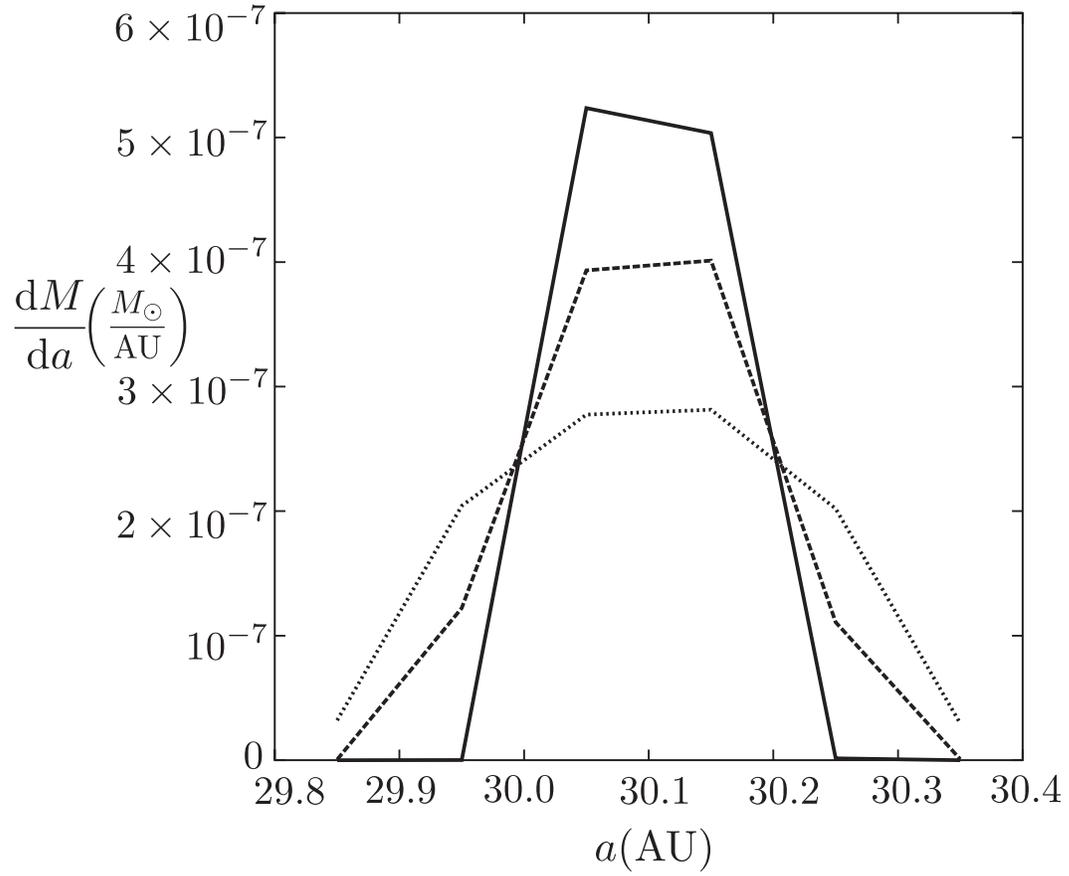}
\end{center}
\caption{Planetesimal disk surface density ${\rm d}M/{\rm d}a(M_\odot/{\rm AU})$
as function of semi major axis.
The solid line is for $t=0$ years,
dashed line is for $t=1000$ years and the dotted line is for $t=10000$ years
in S30e0.05.}
\label{fig.surface_density_0.05}
\end{figure}

Second, in our simulations, the planetesimals tend to spread out
radially, since the width of the initial planetesimal ring is still
rather narrow (figure \ref{fig.run1614_new_eta0.05}).  The surface
density in the radial range of 30.0 - 30.2 AU decreased by roughly a
factor of two from the initial value (figure
\ref{fig.surface_density_0.05}), and the ``average'' density by
roughly a factor of three.  Thus, the total non-binary-induced
collision rate should have decreased by factor between five and ten
(Goodman and Hut 1993).  Therefore, the relative frequency of the
binary-induced collision would be a few times higher than what is
observed in our model, if the radial diffusion of planetesimals are
suppressed.


\section{Summary}
We carried out N-body simulations of planetary growth in the outer
region of the disk in order to study the effect of binaries.
Planetesimals with realistic size distribution are considered.  Our
main findings are summarized as follows.  Binaries are formed in the
outer region of the disk.  They are formed through three-body
encounters.  The random velocities of the planetesimals increase with
time.  Planetesimals with large mass tend to become binary components.
The binaries collide through three-body encounters with third bodies.
Compared to non-binary-induced collisions, binary-induced collisions
prefer massive primaries and very strongly prefer massive secondaries.
As a result, the growth of planetesimals happens mainly through
binary-induced collisions.  We estimate that growth timescale can be
reduced by a factor of five to ten, due to the formation of binaries.

Since the initial planetesimal rings we used are narrow, the
planetesimals tend to spread out radially and the density decreases
from the initial value.  The collision rate and the binary formation
rate decrease as well.  The actual number of the binary formation rate
and binary-induced collision rate could be several times higher if the
reduction of the surface density is prevented.  Hence, the effect of
the planetesimal binary formation onto the accretion rate of
planetesimals should be significant in TNO region.

Timescale for the formation of Uranus and Neptune might be reduced by
this factor, when in situ formation is assumed.
The mechanism we propose will help us to understand the formation of  
outer planets and large TNOs.

\newpage 

{\large{Acknowledgment}}\\ Part of the results is obtained by using
the K computer at the RIKEN Advanced Institute for Computational
Science (Proposal number hpci130026).  This work was supported in part
by MEXT SPIRE Field 5 “The origin of matter and the universe” and
JICFuS.  J.K. thanks Hiroshi Daisaka for letting use his computer
resources, fruitful discussion and encouragements.  We thank the
anonymous referee for the careful and detailed comments which helped
us to improve the paper.


\newpage

\newpage

\newpage

\newpage

\newpage

\newpage

\newpage

\newpage

\newpage

\newpage

\newpage

\newpage

\newpage

\newpage

\newpage

\newpage 
\begin{table}
\begin{center}
\begin{tabular}{c c c c c c c c} 
  \multicolumn{8}{c}{Table 1}\\
  \multicolumn{8}{c}{Model List}\\ 
  \hline
  Run &  $n_{\rm init}$
  & $\eta$
  & $a_{\rm min},a_{\rm max}$(AU)
  & $n_{\rm b}$
  & $n_{\rm b}/n_{\rm init}$ (\%)
  & $n_{\rm col,b}$
  & $n_{\rm col,total}$\\
  \hline
  S30e0.05 & 38283 & 0.05 & 30,30.2 & 72 &0.19 &103  & 243 \\
  \hline
  S30e0.10 & 38283 & 0.10 & 30,30.2 & 13 &0.034& 71 & 149 \\
  \hline
  S30e0.20 & 38283 & 0.20 & 30,30.2 & 9  &0.024& 26 & 66 \\
  \hline
  S30e0.40 & 38283 & 0.40 & 30,30.2 & 6  &0.016& 11 & 28 \\
  \hline
  S10e0.05 & 22000 & 0.05 & 10,10.067 & 4 &0.018&63  & 283 \\
  \hline
  S10e0.10 & 22000 & 0.10 & 10,10.067 & 5  &0.023&26  & 197 \\
  \hline
  S10e0.20 & 22000 & 0.20 & 10,10.067 & 1 & 0.0045&19 & 137\\
  \hline
  S10e0.40 & 22000 & 0.40 & 10,10.067 & 0 & 0 &4  &58 \\
  \hline
  S3e0.05  & 11927 & 0.05 & 3,3.02  & 1 & 0.0084&92  & 450 \\
  \hline
  S3e0.10  & 11927 & 0.10 & 3,3.02  & 0 &0& 23 & 324\\
  \hline
  S3e0.20  & 11927 & 0.20 & 3,3.02  & 1 &0.0084& 9 & 266 \\
  \hline
  S3e0.40  & 11927 & 0.40 & 3,3.02  & 0 &0& 7 & 222 \\
  \hline
  S1e0.05  & 1713  & 0.05 & 1,1.0067  & 0&0 & 20   & 141 \\
  \hline
  S1e0.10  & 1713  & 0.10 & 1,1.0067  & 0& 0 & 8 & 114 \\
  \hline
  S1e0.20  & 1713  & 0.20 & 1,1.0067  & 0&0 & 4 & 97\\
  \hline
  S1e0.40  & 1713  & 0.40 & 1,1.0067  & 0 &0& 0 & 84\\
  \hline
\end{tabular} 
\end{center}
\caption{Table of model list.  $n_{\rm init}$ is the initial number of planetesimals.
Second column $\eta$ is the initial random velocity factor which is explained in section 2.
$a_{\rm min}$ and $a_{\rm max}$
is the minimum value and maximum value of initial disk. $n_{\rm b}$ is the 
number of binaries after $10^4$ years. The percentage expression of the starting
population is the next column.  $n_{\rm col,b}$ is the number of binary-induced
collisions during the simulation.  $n_{\rm col,total}$ is the total number of 
collision.}
\end{table}


\begin{thebibliography}{}
\bibitem [Askatov et al. (2005)]{}
 Askatov S.A., A.L. Ernestine \& D. Farrelly. \ 2005, 
MNRAS, 360, 401-415
%
\bibitem [Chapman et al. (1995)]{}
Chapman et al. \ 1995,
Nature, 374, 783-785.

%
\bibitem [Funato et al. (2004)]{}
Funato, Y., J. Makino, P. Hut, E. Kokubo \& D. Kinoshita. \ 2004,
Nature, 427, 518-520.

%
\bibitem [Goldreich et al. (2002)]{}
Goldreich,P., Y. Lithwick \& R. Sari. \ 2002,
Nature, 420, 643-646.

%
\bibitem [Grundy et al. (2007)]{}
Grundy, W. M. et al. \ 2007,
Icarus, 191, 286-297.

%
\bibitem [Hayashi (1981)]{}
Hayashi, C. \ 1981,
Progress of Theoretical Physics Supplement, No. 70, pp. 35-53.

%
\bibitem [Ida \& Makino (1992)]{}
Ida, S. \& J. Makino. \ 1992.
Icarus, 96, 107-120. 

%
\bibitem [Johnston (2012)]{}
Johnston, W. R. \ 2012.
NASA Planetary Data System, EAR-A-COMPIL-5-BINMP-V5.0

%
\bibitem [Kokubo \& Ida (1996)]{}
Kokubo, E., \& S. Ida. \ 1996.
Icarus, 123,180-191.

%
\bibitem [Kokubo \& Ida (1998)]{}
Kokubo, E., \& S. Ida. \ 1998.
Icarus, 131, 171-178.


\bibitem [Kokubo et al. (1998)]{}
Kokubo, E., K. Yoshinaga \& J. Makino. \ 1998.
MNRAS, 297, 1967-1072.

%
\bibitem [Kominami et al. (2011)]{}
Kominami, J.D., J. Makino \& H. Daisaka \ 2011.
PASJ, 163, 1331-1344.

%
\bibitem [Makino \& Aarseth (1992)]{}
Makino, J., \& S. J. Aarseth \ 1992.
PASJ, 44,141-151.

%
\bibitem [Makino et al. (1998)]{}
Makino, J., T. Fukushige, Y. Funato \& E. Kokubo. \ 1998.
NewA, 3, Issue 7, 411-417.

%
\bibitem [Makino et al. (2007)]{}
Makino, J., K. Hiraki \& M. Inaba \ 2007.
Proceeding of the 2007 ACM/IEEE conference on Supercomputing, 1-11.

%
\bibitem [Marchis et al. (2011)]{}
Marchis, F., J. E. Enriquez, J. P. Emery, J. Berthier, P. Descamps, F. Vachier \ 2011.
Icarus, 213, 252-264.

%
\bibitem [Merline et al. (2000)]{}
Merline, W. J., L. M. Close, C. Dumas, J. C. Shelton, F. Menard, C. R. Chapman, D. C. Slater \ 2000.
American Astronomical Society, DPS Meeting No.32, No13.06; Bulletin of the American Astronomical Society, Vol. 32, p.1017.
%
\bibitem [Nakazawa \& Ida (1988)]{}
Nakazawa K. \& S. Ida \ 1988.
Prog.Theoy.Phys.Suppl., 96, 167-174.

%
\bibitem [Nesvorn${\rm \acute{y}}$ et al. (2010)]{}
Nesvorn${\rm \acute{y}}$, D., A. N. Youdin \& D. C. Richardson \ 2010.
AJ, 140, 785-793.

%
\bibitem [Noll et al. (2008a)]{}
Noll, K. S., W. M. Grundy, E. I. Chiang, J.-L. Margot \& 
S. D. Kern, \ 2008a.
University of Arizona Press, pp. 345-363.

%
\bibitem [Noll et al. (2008b)]{}
Noll, K. S., W. M. Grundy, D. C. Stephens, H. F. Levison \&
S. D. Kern \ 2008b.
Icarus, 194, 758-768.

%
\bibitem [Pravec et al. (2006)]{}
Pravec, P. et al. \ 2006.
Icarus, 181, 63-93.

%
\bibitem [Stephens \& Noll (2006)]{}
Stephens, D.C. \& K. S. Noll \ 2006.
Astron. J., 131, 1142-1148.

%
\bibitem [Walsh \& Richardson (2006)]{}
Walsh, K. J. \& D. C. Richardson \ 2006.
Icarus, 180, 201-216.

%
\bibitem [Walsh et al. (2008)]{}
Walsh, K. J., D. C. Richardson \& P. Michel \ 2008.
Nature, Volume 454, Issue 7201, pp. 188-191.

%
\bibitem [Walsh et al. (2012)]{}
Walsh, K. J., D. C. Richardson \& P. Michel \ 2012.
Icarus, 220, 514-529. 

%
\bibitem [Weidenschilling (2002)]{}
Weidenschilling S.J. \ 2002.
Icarus, 160, 212-215. 

\end{thebibliography}
\end{document}